% MNSAMPLE.TEX
%\vbox to 646pt{} ;tama mahdollistaa kuvien oikean tulostuksen kaksipalstaisessa
%\input epsf

\input psfig.sty 

%
% A sample plain TeX single/two column Monthly Notices article.
%
% v1.5  --- released 25th August 1994 (M. Reed)
% v1.4  --- released 22nd February 1994
% v1.3  --- released  8th December 1992
%
% Copyright Cambridge University Press

% The following line automatically loads the mn macros if you are not
% using a format file.
\ifx\mnmacrosloaded\undefined \input mn\fi

% If your system has the AMS fonts version 2.0 installed, MN.tex can be
% made to use them by uncommenting the line: %\AMStwofontstrue
%
% By doing this, you will be able to obtain upright Greek characters.
% e.g. \umu, \upi etc.  See the section on "Upright Greek characters" in
% this guide for further information.

\newif\ifAMStwofonts
%\AMStwofontstrue

\ifCUPmtplainloaded \else
  \NewTextAlphabet{textbfit} {cmbxti10} {}
  \NewTextAlphabet{textbfss} {cmssbx10} {}
  \NewMathAlphabet{mathbfit} {cmbxti10} {} % for math mode
  \NewMathAlphabet{mathbfss} {cmssbx10} {} %  "   "    "
  \ifAMStwofonts
    \NewSymbolFont{upmath} {eurm10}
    \NewSymbolFont{AMSa} {msam10}
    \NewMathSymbol{\upi}     {0}{upmath}{19}
    \NewMathSymbol{\umu}     {0}{upmath}{16}
    \NewMathSymbol{\upartial}{0}{upmath}{40}
    \NewMathSymbol{\leqslant}{3}{AMSa}{36}
    \NewMathSymbol{\geqslant}{3}{AMSa}{3E}

  \else
    \def\umu{\mu}
    \def\upi{\pi}
    \def\upartial{\partial}
  \fi
\fi

% Marginal adjustments using \pageoffset maybe required when printing
% proofs on a Laserprinter (this is usually not needed).
% Syntax: \pageoffset{ +/- hor. offset}{ +/- vert. offset}
% e.g.    \pageoffset{-3pc}{-4pc}

\pageoffset{-2.5pc}{0pc}

\loadboldmathnames

% \Referee   %  uncomment this for referee mode (double spaced) *******

% \pagerange, \pubyear and \volume are defined at the Journals office and
% not by an author.

%\onecolumn        % enable one column mode ******
% \letters          % for `letters' articles
\pagerange{}    % `letters' articles should use \pagerange{Ln--Ln}
\pubyear{2001}
\volume{}
% \microfiche{}     % for articles with microfiche
% \authorcomment{}  % author comment for footline

\begintopmatter  %  start the two spanning material

\title{Comparison of bar strengths in active and non-active galaxies}
\author{Eija Laurikainen, Heikki Salo and Pertti Rautiainen}
\affiliation{Division of Astronomy, Dep. of Phys. Sci, FIN-90014, Finland}

\shortauthor{E. Laurikainen, H. Salo, P. Rautiainen}
\shorttitle{Bar strengths for active and non-active galaxies}

% \acceptedline is to be defined at the Journals office and not
% by an author.

%\acceptedline{accepted}

email:eija@koivu.oulu.fi

\abstract {Bar strengths are compared between active and non-active galaxies 
for a sample of 43 galaxies, for which bars were identified in the 
near-IR from an original sample of 107 spiral 
galaxies. The bar torques are determined using a new technique 
(Buta $\&$ Block 2001), where tangential forces are calculated 
in the bar region, normalized to the axisymmetric radial force field. 
We use the JHK images of the 2 Micron All Sky Survey (2MASS).
The ellipticities $\epsilon$ of the bars are also estimated 
with an isophotal fitting algorithm and the bar lengths from the
phases of m=2 and m=4 fourier components of density. 
We show a first clear indication that the ellipticity
of a bar, generally used as a measure of the bar strength, is quite well
correlated with the maximum relative tangential force, $Q_b$, in the bar 
region.

We found that nuclear activity occurs preferentially
in those barred early type galaxies, in which the maximal bar torques
are weak ($<Q_b> = 0.21$) and appear
at quite large distances from the galactic center 
when scaled with the radial scale length of the disk ($<r_{Qb}/h> = 1.24$).
For comparison, for the non-active galaxies $<Q_b> = 0.37$ and $<r_{Qb}/h>
= 0.59$. The force maximum in the active late type
galaxies also appears at rather large distances, but
the difference to the non-active galaxies is smaller. 
These results imply that the bulges may be important
for the onset of nuclear activity, but it is not clear why nuclear activity 
appears in some early type galaxies but it is missing in some others.
We also found that for the active early type galaxies the bar length is not 
correlated with bar strength, although a weak correlation appears
for the other barred galaxies studied. Most suprisingly, the 
galaxies with the strongest bars are non-active. } 


\maketitle  %  finish the two spanning material

\section{INTRODUCTION}

It is now widely accepted that non-axisymmetric forces are needed to trigger
nuclear activity by accreting gas to the central regions of galaxies. Primary
bars
are known to be efficient to drag gas in the scale larger than 1 kpc,
but some other mechanism is needed to allow this gas finally to fall into 
the active nucleus, provided for example by secondary bars (Shlosman et al. 
1989). Indeed, star formation 
is found to be enhanced in barred galaxies (Martinet $\&$ Friedli 1997; 
Aquerri 1999), but the connection between bars and nuclear activity 
(AGNs in terms of accretion disks and black holes) is 
not clear observationally. For example, there is no indisputable agreement
showing that the number of bars were larger
for AGNs in comparison to non-active systems. An excess
of bars in AGNs is found by some authors (Knapen et al. 2000, Laine et al. 
2001), while most studies give similar bar fractions for active and 
non-active galaxies 
(McLeod $\&$ Rieke 1995; Moles et al. 1995; Ho et al. 1997; 
Mulchay $\&$ Regan 1997; Hunt $\&$ Malkan 1999, Marquez et al. 2000).
Bars can act as driving forces for the central activity via the strong
inflow of gas in shock regions associated with the rotating bar potential. 
Not suprisingly, the nuclear 
regions of barred galaxies have on average higher concentration of 
molecular gas than 
normal galaxies (Sakamoto et el. 2000, Sheht 2001), which makes 
understandable the found connection between bars and high star formation 
activity.

Star formation activity is found to be correlated with the properties 
of bars, being
enhanced especially in long bars with high ellipticities, usually regarded 
as ``strong'' bars (Martinet $\&$ Friedli 1997; Aquerri 1999). However, 
not all 
long bars have pronounced current star formation activity. On the other hand, 
it has been suggested that Seyferts may even avoid ``strong'' bars 
(Shlosman et al. 2000, Laine et al. 2001), 
but that has been questioned by Marquez et al. 
(2000) who argued that both the lengths and strengths of the primary bars 
are similar for Seyferts and for non-Seyfert galaxies. In all these works 
bar strength is estimated indirectly from the ellipticity of a bar, 
based on the analytical work by Athanassoula (1992a).

The size of a bar is related to the Hubble type so that bars in early type 
systems 
are generally longer than in late type systems, when normalized to the 
galaxy diameter
at 25 magnitude isophote $D_{25}$ (Elmegreen $\&$ Elmegreen 1985, Duval $\&$ 
Monnet 1985, 
Martin 1995, Regan $\&$ Elmegreen 1997). Also, for the bar axial ratio 
Elmegreen $\&$ Elmegreen (1985) find a weak correlation with the Hubble type, 
but that has not been
confirmed by Martin (1995) by a larger sample of galaxies.
Elmegreen $\&$ Elmegreen (1985) also showed that bars in early type galaxies
are generally flat, while those in late type systems are exponential. 
Altogether, the Hubble type is expected to be an important factor in
controlling the properties of bars and probably also for the inflow of gas.

The ellipticity of a bar is not a full description of its 
strength, depending also on the mass of the bar. Morever, the relative
perturbation associated with the bar depends on the central force
field, i.e. the presence of a massive bulge. Therefore bar strengths are 
better 
evaluated by the tangential forces normalized to the total axisymmetric 
force fields, as suggested by Combes $\&$ Sanders (1981) and first applied
for galaxies by Buta $\&$ Block (2001). We use a similar approach
and determine bar torques for 107 spiral galaxies in JHK-bands. Fourier
analysis is used to identify bars, and for those galaxies with
well defined bars, strengths are compared  
between the active and non-active galaxies. We also test how well the bar 
strength and the ellipticity of a bar are correlated.

\section{THE SAMPLE AND THE METHOD}

The original sample consists of those spiral galaxies with $B_T < 12.5$ mag, 
$cz < 2500 \ km \ sec^{-1}$ and $i < 67^0$ in the Third Reference 
Cataloque of Bright Galaxies (hereafter RC3), for which high quality 
images 
were available in the 2 Micron All Sky Survey (hereafter 2MASS). By i
we denote the inclination of the galactic disk.
Additionally some of the weakest objects were eliminated so that the  
number of galaxies in the sample was 107.    
To active galaxies we included Seyferts, LINERs and HII-galaxies, for which
the spectral classifications were taken from the NASA/IPAC Extragalactic
Database (NED), where the latest spectral classifications are available.
Altogether, 53 of the galaxies show nuclear activity and 31 are barred
classified as SB in RC3. 
The image resolution was 1''/pixel and the H-images 
were generally deeper than the images in the J or K-bands. 
The selection effects of the sample are discussed by Laurikainen $\&$ Salo 
(2001). We found that the frequency of bar identifications 
rapidly decreased when the inclination of the disk is larger than $50^0$.
Also, the non-active galaxies appeared to be on the average somewhat brighter 
than the active galaxies. Considering that the absolute brightnesses of the 
galaxies correlate with the bar lengths, on the average somewhat longer
bars are thus selected for the active galaxies. However, we also estimated
that these biases do not affect our conclusions. 

The distribution of optical morphological types for active 
galaxies in our sample is quite similar to that found previously for 
Seyferts in larger samples of galaxies, the 
active galaxies being shifted towards earlier Hubble types.
The peak appears for Sab types, which is between the mean morphological
types for Seyfert 1 (Sa) and Seyfert 2 (Sb) galaxies by Malkan et al.
(1998). If we exclude HII-galaxies, 
the active systems are even more clearly concentrated to early Hubble types.
In the following the galaxies will be 
divided to early (SO/a, Sa, Sab), late (Sbc, Sc, Scd) and very late types 
(later than Scd), based on the classification in RC3. The omission of the 7 
latest type systems from the category of late type galaxies
is justified by the fact that for types later than Scd the bulge-to-disk 
ratio does not follow
the general decreasing tendency from early to later types 
in the Hubble sequence (de Jong 1996), which phenomenon has been 
theoretically discussed by Noguchi (2000). 

The galaxies were divided to barred and non-barred systems so that the 
identification of a bar
was based on the near-IR images, which identification is independent
of that given in RC3 (SA, SAB, SB). Thus from hereon our term 'barred'
is not the same as 'SB'. 
Instead of identifying bars 
visually we used fourier techniques. The presence of a bar was confirmed 
if the phases of the m=2 and m=4 fourier components of density were 
maintained nearly constant in the
bar region. Additionally, it was required that the m=2 component had a 
high amplitude. 
The length of the region where the phase was maintained constant also
determined the length of the bar. The number of 43 barred galaxies found in 
the near-IR clearly exceeds the number of barred galaxies in optical, in 
agreement with
earlier studies based on near-IR images (Block $\&$ Wainscoat 1991, 
Knapen $\&$ Schlosman $\&$ Peletier
2000, Eskridge et al. 1999). Also, all galaxies classified as SB in RC3,
appeared to be barred in the near-IR. In principle our method of calculating
bar strengths does not require any preidentification of a bar, but here we
wanted to concentrate only to clear cases of bars. 
 
Bar strengths were determined by transforming the light distributions
into potentials and deriving the maximum ratios of the tangential forces
relative to the radial forces. This approach was first suggested by Combes
$\&$ Sanders (1981), but has been applied for galaxies only recently
by Buta $\&$ Block (2001), who used the potential evaluation method by
Quillen et al. (1994) to the force calculation. In order to obtain a 
single measure for the strength we use $Q_b$, which is the maximum of $Q_T$ 
in the bar region (as in Buta $\&$ Block 2001). The distance where 
the maximum $Q_T$ occurs is denoted by $r_{Qb}$. For the  distances we used
the measurements by Tully (1988). 

Our method of force calculation is not completely identical with that by 
Buta $\&$ Block, 
the full description of it being presented by Laurikainen $\&$ Salo (2001).
For example, due to the limited resolution of the 2MASS images the 
gravitational potential was not calculated with cartesian integration
from the original images.
Instead the images were first ``smoothed'' by calculating
the azimuthal fourier decompositions of the surface densities in 
different radial
zones similarly as done in Salo et al. (1999). In the calculations the even 
Fourier modes from 2 to 10, characteristics for bars (see Ohta 1996) 
were included. Also, instead of using the vertical scale height of the 
Milkey Way as done in Buta $\&$ Block, it was taken to be a certain 
fraction of the radial scale length of the disk.

The mass density in vertical direction was approximated by an
exponential model. Following de Grijs (1998) we used 
$h/h_z$=2.5 for the early-type galaxies and $h/h_z$=4.5 for the late type
systems. The exponential
scale lengths were taken from the catalogue by Baggett et al. (1998) when
available, and otherwise they were estimated from the 
2MASS images by us. For a few cases where $h$ was not given  
by Baggett et al. and could not
be determined from the 2MASS images either, the scale height 
of the Milky Way ($h_z$=325 pc) was used in the force calculation. These 
galaxies were omitted  from those plots,
where the bar parameters are scaled to $h$. For
three of the galaxies, pgc 10266, pgc 15821 and pgc 40097 the scale length 
given by Baggett
et al. was considered to be unrealistic. For two of them $h$ was rather a 
measure
for the brightness slope in the bulge region, whereas for pgc 10266 the given 
scale length represented the outermost very shallow part of the 
disk, while we are interested in the disk under the bar. Also for 
these three
galaxies the assigned scale lengths were measured from the 2MASS images.

The ellipticity of a bar, generally used as a measure of the bar strength, 
is an approximation of the true bar strength. Therefore, for comparison 
the maximum ellipticities
were also determined by fitting ellipses to the isophotes of the surface 
brightnesses, as described
by Laurikainen $\&$ Salo (2000). The ellipticity of a bar was taken to be 
that of the smallest deprojected minor-to-major axis ratio 
$\epsilon = 1-(b/a)_{min}$. 
Our measurements are in good agreement with those by Laine et al. (2001)
for 6 barred galaxies common in our samples, both regarding $\epsilon$ and the 
semimajor axis of the most elongated isophote ($r_{\epsilon}$) (see Fig. 1).
The mean values of $Q_b$, $\epsilon$ and their radial distances for the 
different
subsamples are shown in Table 1. The errors in the table denote the 
sample standard deviations.
For $Q_b$ the largest source of uncertainty is in fact due to the 
uncertainty in the vertical
scale height, which for example for Sc-galaxies varies by $h/h_z$ = 2.5 
to 5.5 (de Grijs 1998).
This corresponds to an uncertainty of about 15 $\%$ in $Q_b$. 
Notice that the uncertainty in $h_z$ does not affect $r_{Qb}$. 

\section{STRENGTHS OF THE NON-AXISYMMETRIC FORCES}

In the following non-axisymmetric forces are compared between the active and 
nonactive barred galaxies. Since the measurements in the J, H and K-bands 
give very similar results, only those related to H-images are reported.
Bar strengths
between the early and late type galaxies are also compared. This is important
because the Hubble type to some extent measures the 
bulge-to-disk ratio (B/D) for a galaxy and the bulge might be an important 
factor in controlling the properties of bars and nuclear activity.
The B/D-ratio varies along the Hubble sequence similarly in the optical
and in the near-IR (de Jong 1996).

Before doing any such comparisons between different galaxy subsamples 
it is interesting to verify how
well the maximum isophotal ellipticy $\epsilon$ of a bar
and the maximal non-axisymmetric force $Q_b$, are correlated. Indeed, 
we found rather tight correlation between these two parameters (see Fig. 2).
We show here b/a = 1 - $\epsilon$, to make it
easier to compare with the similar plot by Block et al. (2001).
Fig. 2 also uses different symbols for various subsamples, indicating 
that there is a small difference between
the early and late type galaxies in the diagram, being largely
due to the larger vertical scale height used for the early type galaxies:
if the same $h/h_z$ is used for all galaxies the difference would disappear.
This correlation is the first direct observational confirmation
showing that the bar axial ratio is indeed a good measure of the bar strength.
However, it is worth noticing that especially when the ellipticity of a 
bar is high, even a small change in the elongation of a bar results to 
a large change in the non-axisymmetric force. Therefore, $Q_b$ is a more
sensitive measure of the bar strength even for strong bars. 

A correlation between $Q_b$ and b/a was found also by Block et al. (2001), 
but the dispersion was two or three times larger than in our similar diagram.
Based on this scatter they suggested that
apparently bars with significant ellipticities may be either strong, weak
or intermediate as far as the bar torques are concerned. However, on the basis
of our result this
probably is not the case. We rather suppose that the most important reason
for the large scatter was the high uncertainty
in the b/a values by Martin (1995) that Block et al. used.  
Martin estimated the uncertainties of $\epsilon$ to be 
about 20 
$\%$. The uncertainties are large, because the lengths of the major and
minor axis were estimated visually without any isophotal-fitting routine. 
Also, blue
photographic plates were used which might be another cause for the
large scatter: the near-infrared images
that we use are better expected to trace the true mass distribution 
in the bar region than the blue 
images. The fact that Block et al. included also the
SAB-galaxies to their diagram hardly explains their scatter, because 
it was large also in the region with $Q_b > 0.3$, where only 
SB-galaxies appeared. 

When all barred galaxies are considered (see Fig. 3, Table 1) it seems that 
active galaxies
might have on the average smaller non-axisymmetric forces than the
non-active systems ($<Q_b>$=0.27 v.s. 0.37). However, while applying 
the Kolmogorov-Smirnov
test (KS-test) this difference is only marginally significant: the 
probability that the samples are drawn from similar
populations is p=0.10. 
A similar result, but with no statistical
significance (p=0.37) was obtained for the ellipticities of bars. 
Our result for the ellipticities is in accordance with Shlosman et al. 
(2000) and by Laine et al. (2001). 
In our preliminary study (Laurikainen et al. 2001) it was 
argued that the difference in $Q_b$ between the active and non-active galaxies
is statistically significant with p=0.05.
The reason for the slightly lower statistical confidence level in the current 
study is that here the sample has been somewhat reduced by lowering the upper
inclination limit. Also,
bar strengths for a few galaxies have been remeasured using more proper 
scale lengths of the disks. 

\begintable*{1}
\caption{{\bf Table 1.} Mean maximum tangential forces, maximum ellipticities 
and the locations of the maxima, as well as bar lengths for barred galaxies.
The errors denote the standard deviations.}
\halign{%
\rm#\hfil&\qquad\rm#\hfil&\qquad\rm\hfil#&\qquad\rm\hfil
#&\qquad\rm\hfil#&\qquad\rm\hfil#&\qquad\rm#\hfil
&\qquad\rm\hfil#&\qquad\rm#\hfil&\qquad\hfil\rm#\cr
 & N & $<Q_b>$ & $<r_{Qb}/h>$&$<r_{bar}/h>$ & $<\epsilon>$ & $<r_{\epsilon}/h>$ & & & \cr
\noalign{\vskip 10pt}

act          &21 & 0.27$\pm$0.13 &1.15$\pm$0.61 &1.85$\pm$0.79 &0.57$\pm$0.11 &1.40$\pm$0.77 &  &  & \cr
non-act      &22 & 0.37$\pm$0.23 &0.59$\pm$0.30 &1.26$\pm$0.52 &0.60$\pm$0.14 &0.83$\pm$0.44 &  &  & \cr
early        &19 & 0.25$\pm$0.13 &1.04$\pm$0.61 &1.80$\pm$0.82 &0.56$\pm$0.12 &1.31$\pm$0.72 &  &  & \cr
late         &21 & 0.38$\pm$0.23 &0.73$\pm$0.46 &1.39$\pm$0.60 &0.61$\pm$0.12 &1.00$\pm$0.62 &  &  & \cr
act-early    &14 & 0.21$\pm$0.07 &1.24$\pm$0.58 &1.99$\pm$0.83 &0.53$\pm$0.09 &1.50$\pm$0.73 &  &  & \cr
act-late     & 6 & 0.40$\pm$0.13 &0.92$\pm$0.73 &1.58$\pm$0.76 &0.67$\pm$0.06 &1.11$\pm$0.90 &  &  & \cr
nonact-early & 5 & 0.35$\pm$0.19 &0.51$\pm$0.28 &1.27$\pm$0.54 &0.63$\pm$0.17 &0.77$\pm$0.33 &  &  & \cr
nonact-late  &17 & 0.38$\pm$0.25 &0.61$\pm$0.32 &1.25$\pm$0.53 &0.60$\pm$0.13 &0.85$\pm$0.49 &  &  & \cr
 & & & & & & & & & \cr
}
%\tabletext{\noindent $^a$Observed by {\it IRAS}.}
\endtable

However, statistically significant differences appear between the active 
and non-active galaxies
for the locations of the maximum tangential forces, $r_{Qb}/h$, 
and in the radial distances of the maximum
ellipticities, $r_{\epsilon}/h$. These maxima appear in much
larger distances from the galactic center for the active than for the
non-active galaxies (Fig. 3, Table 1). The probablilities that the compared
samples are drawn from similar populations are
p=0.0008 and 0.007 for $r_{Qb}/h$ and $r_{\epsilon}/h$, respectively. As 
the $r_{Qb}/h$-value is quite sensitive
to the scale length of the disk, we normalized $r_{Qb}$ also to the
diameter at the contour level of 25 $ mag \ arcsec^{-1}$
$D_{25}$, but that did not affect the conclusions or the level of 
the statistical significances in the comparisons.   
The difference in the barred properties between the active and non-active 
galaxies is even more illustrative while correlating $Q_b$ with
$r_{Qb}/h$, as shown in Fig. 4. It is remarkable that almost all galaxies with
$r_{Qb}/h > 1$ are active, while for the non-active
galaxies $r_{Qb}/h$ is rather small even for the strongest bars.
However, an important point to stress here is that due to their morphological 
distributions the active galaxies largely follow the distribution of
early-type galaxies, whereas the non-active galaxies behave much like 
the late-type systems. Thus the difference we find might simply indicate
a difference between the early and late-type spirals. However, the
connection between the activity and the Hubble type is not that
straightforward as will be discussed in the following.

The radial $Q_T$-profiles for the individual barred galaxies  
are presented in Fig. 5, showing separately the active and 
the non-active early and late type galaxies. While comparing the average 
bar strengths we can see that in fact only 
the early type active galaxies have noticably smaller bar strengths, with 
the mean $<Q_b>$=0.21.
All the other subsamples, such as the non-active early type galaxies 
and the late type 
systems have larger $<Q_b>$ values between 0.35-0.40 (see Table 1).
The KS-test shows that the probability that the samples of active and 
non-active early type galaxies are drawn from similar populations is
p=0.04 and p=0.02 for $Q_b$ and $\epsilon$, respectively, which means 
that these differences are statistically significant.
 
It is also clear that the distances of the ellipticity maxima and the 
maximal tangential 
forces are largest for the active early type galaxies 
with $<r_{Qb}/h>=1.24$ and $<r_{\epsilon}/h>=1.50$, respectively (see Fig. 5
and Table 1).
For the active late type systems these parameters show somewhat
lower values with $<r_{Qb}/h>$=0.92 and $<r_{\epsilon}/h>$=1.11, but they 
are still higher than for the non-active galaxies with
$<r_{Qb}/h >$ = 0.51-0.61 and $<r_{\epsilon}/h>$=0.77-0.85. 
These results show that the bulges alone cannot explain why especially 
the active early type galaxies have small $Q_b$ and large $r_{Qb}$-values.

\section{BAR LENGTHS}

The lengths of the bars were estimated both by fourier techniques ($r_{bar}$)
and by the maximal ellipticities of bars ($r_{\epsilon}$). In Laurikainen 
$\&$ Salo (2001) the absolute bar length in kiloparsecs was found to 
correlate with the 
absolute brightness of the galaxy, in agreement with Kormendy (1979),
but while scaling the bar length to the scale length 
of the disk, the correlation disappeared. Therefore,
when using the scaled bar lengths ($r_{bar}/h$), there is no need to worry
about possible magnitude biases in the compared samples.
We found that $r_{bar}$ correlates with $r_{\epsilon}$ for all Hubble 
types and activity classes (see Fig. 6, lower panel). However, 
$r_{\epsilon}$ gives systematically
shorter bar lengths, which means that $r_{\epsilon}$ is not a very reliable
measure of the bar length, often underestimating the true bar length.
In fact, in many N-body simulations the bar ellipticity can decrease 
considerably 
before the actual end of the bar (Rautiainen $\&$ Salo 1999). The 
distance $r_{\epsilon}$ is also
correlated with $r_{Qb}$, but in such a way that the force maximum appeared
systematically at shorter distances than the ellipticity maximum
(see Fig. 6, upper panel). 

We confirm the earlier result by Elmegreen $\&$ Elmegreen (1985),
Martin (1995) and Regan $\&$ Elmegreen (1997) showing that early type 
galaxies have on
average longer bars than late type systems (see Table 1), but the 
difference we find is
smaller than suggested by Martin. In fact, the samples by Martin and 
Regan $\&$ Elmegreen
are not very representative for early type spirals: Regan $\&$ Elmegreen have
only 1 early type galaxy among 23 galaxies, and Martin has 5 galaxies among 
136, classified as 
SO/a, Sa or Sab. On the other hand, our result is in accordance with the 
study by Elmegreen 
$\&$ Elmegreen (1985), based on a sample of 99 galaxies, which covers well
the whole range of the Hubble sequence for spiral galaxies. As in 
our work, they also find larger dispersion in bar lengths for the early 
than for the late type
galaxies, and especially some SO/a galaxies in their sample have very
short bars. 

When the whole sample of barred galaxies is studied, bar strength $Q_b$ 
(or $\epsilon$) 
definitely does not correlate
with bar length $r_{bar}$, which is shown in Fig. 7. This is the 
case both 
when the bar length was given in kiloparsecs, or scaled to the scale length
of the disk. Therefore, bar length cannot be considered as an indice of  
bar strength: short bars can have either strong or weak tangential forces.
However, if only the late type or the non-active early type galaxies are 
considered, $r_{bar}$ seems to slightly increase with the increasing $Q_b$, 
which is in 
accordance with Martin (1995), whose sample consisted mainly of late type 
spirals. The suprising thing here is that the active early type galaxies
have rather long bars, eventhough their strengths are only weak or moderate.

Like bar strengths, also bar lengths for the active and non-active galaxies
in our samples are associated to the 
Hubble type: while bars of active galaxies have rather similar lengths
with the early type galaxies, bars in non-active 
systems largely follow the length distribution of late type galaxies 
(see Fig. 7).
The longest (and at the same time weakest) bars belong to 
the active early type galaxies, whereas bars in the late type galaxies
are shorter (see also Table 1). This is a result that 
should be understood also
theoretically: why in the bulge dominated galaxies where the bulge probably 
stabilizes the bar region thus reducing the bar strength, 
bars are at the same time very long? This will be discussed in Chapter 7.

\section{Bar ellipticity and tangential forces: analytical toy models}

In order to gain understanding of the found dependence between 
$Q_b$ and $\epsilon$ some simple analytical force models
were constructed. We use a model potential which consists of a
spherical Plummer bulge, an axially symmetric exponential disk and a
non-axisymmetric bar, represented by a prolate Ferrers-ellipsoid.  The
bulge and disk are characterized by their mass, $M_{bulge}$ and
$M_{disk}$, and by the bulge radius $R_{bulge}$ and the disc exponential
scale-lenght $h$. In all cases the ratio $R_{bulge}/h=1/5$.
The Ferrers-ellipsoid has a density function
\vskip 0.3cm

$\rho = \rho_0 (1- g^2)^n  \ \   {\rm if} \ \ g<1$,
\vskip 0.3cm

$\rho = 0  \ \ \  {\rm  if} \ \ g>1$,
\vskip 0.3cm

\noindent with $g^2 = x^2 / a^2 + (y^2 + z^2)/b^2$, where $a$ and $b$ 
stand for the bar major and minor axis and $\rho_0$ for its central
density, connected to the total bar mass by $M_{bar}=2^{2n+3}
\Gamma(n+1)\Gamma(n+2)/\Gamma(2n+4) \rho_0 \pi ab^2$ (Athanassoula
1983). The values $n=0,1,2$ where considered, $n=0$
representing a bar with constant density, while in the case $n>0$ the bar
is more centrally condenced. The potential corresponding
to the bar density distribution was constructed with the formulas
given by Pfenniger (1984), and the radial and
tangential force components in the equatorial plane $z=0$ were
obtained by numerical differentiation. As adviced by Pfenniger
(1984), the forces were checked by observing that Poisson equation
was satisfied. The mean radial force due to bar was obtained from the
average over different azimuthal directions. In all models the lenght
of the bar major-axis was fixed to $a=2h$.

In the construction of $Q_T$-profiles, two basic
rotation curve models were studied, differing in the amount of bulge
mass with respect to the combined bar + disk mass, having
$M_{bulge}/(M_{bar}+M_{disk})=0.3$ and $0$. The combined bar+disk mean
radial force was fixed to that due to a bar with $b/a=0.5$. Thus in the
case of different $b/a$ ratio the axisymmetric disk and bulge
actually deviate from those defined above, in a manner that would
yield the desired total bulge+disk+bar radial force. For very
elongated bars the implied radial force due to the bar alone would 
in some cases exceed the total radial force: these unrealistic models 
were excluded.

In Figs. 8 and 9 the implied $Q_b$ and $r_{Qb}/a$ distances are studied
as a function of $b/a$ ratio, for different values of
$M_{bar}/(M_{bar}+M_{disc})$ and $n$. Also shown are the rotation curves
corresponding to the mean radial forces, with slight differences
caused by different $M_{bar}/(M_{bar}+M_{disc})$ and $n$, as well as
examples of $Q_T$-profiles (for $b/a=0.5$). The former figure
corresponds to the model including the bulge component, being
characterized by a steeply rising inner rotation curve, while in
the latter figure the bulge is omitted, leading to a more shallow rise
of the rotation curve. With our adopted model parameters the rising
portions have lenghts of about $0.1 a$ and $0.5a$.
Very roughly, these two models could be interpreted as
representing those of early type spirals (dominant bulge) and late
type spirals (weak bulge). Also shown in the plots are the measured
values of $Q_b$ and $r_{Qb}/r_{bar}$  for the barred galaxies in our sample.

Inspite of its simplicity, the Ferrers-bar model seems to account
fairly well for the general trend of $Q_b$ vs. $b/a$, suggesting the 
possibility that the observed scatter
arises due to the different bar mass fractions (Fig. 8 upper row). 
Interestingly, the value of adopted $n$ changes $Q_b$ only very
little, suggesting that also more realistic bar profiles might lead to
very similar $Q_b$ for a given bar mass (Fig. 8 lower row). On the other hand,
the location where maximum tangential force is obtained depends
sensitively on $n$, more concentrated bar models (larger $n$) leading to
smaller $r_{Qb}/a$. In the case with no bulge (Fig. 9) the $Q_b$'s are
naturally somewhat increased for a given $M_{bar}/(M_{bar}+M_{disc})$, due
to weaker total radial force. The increased lenght of the rising
portion of the rotation curve (= portion of much reduced mean radial
force) is also visible in $r_{Qb}/a$, where models with very
elongated bars tend to have maxima shifted to very small distances.
Altogether, the scatter in observed $r_{Qb}/r_{bar}$ vs. $b/a$
is also rather nicely accounted for.

\section{COMPARISON OF SEYFERTS, LINERs AND HII-GALAXIES}

We next compare the properties of bars in Seyferts, LINERSs and
HII-galaxies. On the basis of the previous work by Martinet
$\&$ Friedli (1997), 
HII-galaxies typically reside in long bars, while 
bars in Seyferts might have on the average similar lengths as 
non-active galaxies (Marquez et al. 2000). Shlosman et al. (2000) and
Laine et al. (2001) found some evidence that
Seyferts might miss bars with large ellipticities, but that has been
contradicted by Marquez et al., who concentrated only on isolated
galaxies.

We showed in Chapter 4 that bar length $r_{bar}$ is weakly correlated with  
bar strength $Q_b$, but only for the late-type galaxies or for the non-active
early-type spirals (see Fig. 7). A similar 
plot is shown for Seyferts, LINERs and HII-galaxies separately in Fig. 10. 
The interesting thing here is that the galaxies with the longest bars
($r_{bar} > 6$ kpc) 
are either type 1-1.5 Seyferts (3 galaxies) or intermediate types between
LINERs and Seyferts 2 galaxies (1 galaxy), while for one of them the Seyfert 
type
is not known. On the other hand, we have no identification of type 1-1.5
Seyferts among the Seyferts with shorter bars ($r_{bar} = 1.5 - 5$ kpc). 
One of them is identified as
Seyfert 2, although for three of them the Seyfert type is not known. All the
type 1-1.5 Seyferts here have early Hubble types, whereas the HII-galaxies
are mostly late type systems. 

Seyfert 2 galaxies generally have strong circumnuclear 
starbursts or may in some cases even have nuclear starbursts, so that in that
sense they can be more closely associated with HII-galaxies. Or at least 
the true 
active nuclei may be overshadowed by strong star formation events.
Also, as LINERs represent 
the lower-level nuclear activity, probably induced by a shock-heating
mechanism, it is possible that the true active nuclei in terms of 
black holes and accretion disks are only of the type 1-1.5 Seyferts. Therefore,
if Sy1-1.5 galaxies could 
really be distinquished from LINERs, type 2 Seyferts and of 
HII-galaxies by their long, but relatively weak 
bars that would probably be a new helpful piece of knowledge when
discussing the formation and evolution of active galactic nuclei.
However, our result is only of a preliminary nature and should be confirmed
using a much larger sample of galaxies.

\section{DISCUSSION}

The connection between bars and nuclear activity has been a long standing
debate since the first efforts by Noguchi (1988) and Shlosman et al.
(1989). They showed that non-axisymmetries in the background
gravitational potential, e.g. stellar bars, intrinsic or induced
by interactions, can cause
redistribution of mass in galactic disks in such a way that it may help to
ignite formation of an active
nucleus. However, this connection has turned out to be extremely difficult to
prove both theoretically and observationally. 

From 60 $\%$ to 75 $\%$ of all galaxies have bars, and even a larger
number of them are barred if minibars are also
included (Regan $\&$ Mulchaey 1999, Martini $\&$ Pogge, 1999, Eskridge $\&$
et al. 1999). There is no 
indisputable agreement showing that Seyferts really
have more bars than the non-active galaxies, and it is still a puzzle 
why nuclear activity appears in some barred galaxies, but does not exist in 
most of them. Of course, both for a nuclear starburst and for fueling an active
nucleus, a minimum amount of gas is needed. This condition seems to be
well fullfilled in many barred galaxies, because they have 
on the average three times higher nuclear molecular gas surface densities than 
the unbarred
galaxies (Sheth 2001). Therefore it is clear that enhanced gas density in 
the bar region 
is not a sufficient condition for the onset of nuclear activity. 

The gas flow associated with the non-axisymmetric bar potential occurs both
inward and outward, depending on the position with respect to the bar major
axis. In the seminal papers by Athanassoula (1992a, b) it was shown that
shock regions are accompanied by a strong inflow, so that net inflow becomes 
possible if high density shocks are present. The form of shocks, and the
strength of connected inflow depends crucially on the main orbital families
of the potential: the eccentricity of the major-axis, bar-supporting $x_1$
family, and the presence and extent of perpendicular $x_2$ orbits, associated 
with the ILR. Especially, off-axis shocks require the presence of $x_2$
orbits. The morphology of the shock features in Athanassoula's models
explains well the observed shapes of dust lanes in barred galaxies, placing
strict limits for the relation between the bar extent and its corotation 
radius for bulge dominated galaxies. In general, more massive and more
elongated bars lead to stronger inflow rate, due to larger density contrast
between shock and non-shock regions. Later studies have confirmed the 
robustness of these results in terms of dependence on various numerical
methods and model parameters (Patsis $\&$ Athanassoula 2000). However, 
the shock morphology and inflow is to some degree sensitive to the effective
sound speed of gas, larger random motions favouring on-axis shocks and
larger inflow (Englmaier et al. 1997, Patsis et al. 2000).

The large fraction of active systems among our early type barred galaxies
(about 3/4) is in accordance with these shock models: the potential 
perturbation accompanied by $Q_b > 0.15$ or $\epsilon > 0.4$
is well in the range expected to cause a substantial inflow of gas. The fact
that we find a smaller fraction of active systems among the late type 
barred spirals (1/4) is also as expected: reducing the central concentration
of the galaxy potential (moving from early to late type systems) first 
limits the extent of $x_2$ orbits and then makes them to disappear
altogether, which according to Athanassoula (1992b) replaces the strong 
off-axis
shocks by the weaker on-axis shocks and finally makes the shocks to 
disappear. In general, the active systems among our late type barred galaxies
all represent fairly large perturbation 
($Q_b  >  0.3$ or $\epsilon > 0.65$), although there are several 
non-active galaxies with similar perturbation strengths. 

Indeed, some important questions arise from our measurements. For example, why
the average bar strength is smaller for the active than for the non-active
galaxies of a similar morphological type (early type galaxies)? And also,
why there is no correlation between bar strength and length for the
active early type galaxies? Most suprising is our finding that the average
perturbation in the non-active early type galaxies is much higher than 
that for the active early type galaxies. Especially, our sample contains 
4 non-active early type 
galaxies with $Q_b > 0.4$ ($\epsilon > 0.63$), for which high 
inflow rate would be expected (see Fig. 4). Perhaps these systems 
represent a case 
where the previous inflow has been so efficient that the fuel available for 
active nucleus has already been consumed. Or perhaps too strong bars in
general are not favourable for supporting nuclear activity: only two of our
active galaxies have a perturbation in the above range, but for them $r_{Qb}/h$
is very large. 

In general, the
moderately weak bar potential favours the inflow of gas toward a nuclear ring 
connected to ILR, provided that the bar pattern speed is not so high that
ILR is completely absent. Nuclear ring represents in itself a rather stable
configuration, which however, might become susceptible to further dynamical
instabilities via gradual build-up of material, as in the original minibar
scenario by Shlosman et al. (1989): these additional mechanisms would
then be responsible for the actual AGN activity. On the other hand, according
to the models by Athanassoula (1992 b) very strong bars (massive or highly
elongated) lead to the disappearance of $x_2$ orbits. Consequently, instead of 
accumulating to nuclear rings the gas flows directly toward center, which
perhaps is not an ideal condition for the further feeding mechanisms to 
operate. The fact that
the active early type galaxies were found to have force maxima at rather
large distances also supports the presence of $x_2$-orbits: these bars probably
have fairly flat density profiles, resembling the homogenous bar models
in Athanassoula (1992b) with substantial extent of $x_2$ orbits. Also, 
their bars were found to be fairly long, suggesting slow rotation and thus the 
presence of ILR. 

The formation and evolution of bars is a complicated process, in which 
secular evolution both in terms of isolated evolution and galaxy interactions
might play an important role, thus changing the barred properties 
or even the Hubble type. 
In general, for weak bars
to develope the dynamical instabilities must be rather small in the inner
regions of the disks. That kind of conditions can be produced in simulations
for example by increasing the stellar velocity dispersion so that the 
Toomre parameter is 2 $<$ Q $<$ 3 (Athanassoula 1992a, Rautiainen $\&$ Salo 
2000), by cold and clumpy
gas in the disk (Shlosman $\&$ Noguchi 1993) or by central mass 
concentrations like compact bulges, nuclear star clusters or supermassive 
black holes (Hasan $\&$ Norman 1990, Hasan et al. 1993, Norman et al. 1996).
That kind of compact structures lead to a subsequent weakening of the 
bar and finally even to its dissolution.

Compact bulges can be formed for example in the evolutive processes of bars,
where thickening of the inner particle distribution occurs when the bar
dissolves, as discussed by Norman et al. (1996). This dense
mass concentration gradually reduces the volume of phase space accessible
to regular, bar-supporting orbits of $x_1$ family. In some of
their models the bar was only weakened during the formation of the new
bulge, but in some cases it was completely destroyed. This kind 
of secular evolution gives one possible explanation why bars can be either 
strong or weak among galaxies with large bulges (active and non-active 
early-type galaxies), but in this scenario it is less clear why the activity 
should appear especially  
in the more weakly barred galaxies, e.g. in those where the bulges are
suggested to be relics of earlier bars. Especially, what mechanism 
would in that case feed the nucleus?
Alternatively, the non-active early type galaxies (with large bulges) might
also be candidates for the relics of this kind of secular evolution of bars.
What we could do is for example to look at the morphological types of 
bars in more detail. For example, in the models by Norman et al. (1996)
the newly formed bulges never
have boxy shapes. Also worth looking at are the two different bar 
components
that somethimes appear simultaneously in galaxies, the thick bar component 
composed of a warm stellar population, and the thin spindle made of
cool population (Block et al. 2001). These two bar components
are possibly formed at different epochs in the life of the galaxy and
therefore might be indices of secular evolution.

There is some evidence that galaxy interactions may also
play an important role in the formation of bars (Noguchi 1988, 1996, 
Salo 1991).
In the models by Noguchi (1996) strong bars can develop 
after the pericenter passage, the formed bars being long lasting with little 
changes in their strengths or lengths. The intensity profiles of bars
produced in tidal processes are typically flat, while those produced without
any external triggering are more rapidly declining. This fits to the
observations by Elmegreen $\&$ Elmegreen (1985) who showed that bars
in early type galaxies are often flat, whereas in the Hubble
types of Sc or later bars are generally exponential. According to Noguchi
the weak response in the galactic disk to the bar instability is ascribed
to highly dissipative gas, which effectively stabilizes the stellar disks
by creating large stellar clumps leading to effective heating of the disk
(Shlosman $\&$ Noguchi 1993). In this
scenario the exponential bars, contrary to the flat bars, are not created
from the disk being rather bulge components deformed by the bar instability
(Noguchi 2000). 

The environmental study of normal galaxies by Elme\-green et al. (1990) gives
some support to the interaction scenario. They showed that among the early 
Hubble types the fraction of barred galaxies
is twice higher in binaries than in field galaxies or galaxies in groups, 
suggesting
a strong link between close interactions and flat bars. Therefore one 
would expect that Seyfert galaxies, which appear preferentially in
early type spirals, would also have frequently bars and appear in
crowded galaxy environments. Indeed, Seyferts
may have more bars than the non-active galaxies, but contrary opinions also
appear. However, it is promising that the frequency 
of Seyfert activity is clearly increased only in the barred early type 
galaxies (Laurikainen $\&$ Salo 2001),
which indicates that after all, large scale bars might somehow be 
controlling the nuclear activity.
Concerning the galaxy environments there are 
at least two observational results showing that galaxy
interactions should not be forgotten while discussing Seyfert activity. 
Namely, Seyferts avoid strongly disrupted interacting systems 
(Keel et al. 1985, Bushouse 1987), which is in the same line with our result 
showing that active galaxies generally have rather small non-axisymmetric 
forces in the inner disks. And secondly, Seyfert 2 
galaxies appear more frequently in interacting systems (Laurikainen $\&$ 
Salo 1995, de Robertis et al. 1998, Dultzin-Hazyan et al. 1999) and have on 
the average
more companions than the type 1 Seyferts or the non-active galaxies 
(Laurikainen $\&$ Salo 1995). 

In order to dissentangle the full dynamical
stages of bars in the galaxies studied here it would be important 
to compare the length and strength properties with the high resolution 
gas kinematical observations available for nearby galaxies. That would for 
example highlight the connection between bar strength and gas inflow 
and possibly also give some perspective to the evolutive
scenarios of bars. It would also be important to
understand in more detail to which extent and how the bulges may control
the barred properties and consequently the nuclear activity in galaxies.

\section{CONCLUSIONS}

We have compared bar strengths in active and non-active galaxies for
a sample of 43 barred galaxies. In order to consider only the clear cases, 
identification of a bar was done in the near-IR by fourier techniques
from an original sample of 107 disk galaxies.
Bar strengths were estimated by a new method (Buta $\&$ Block 2001)
calculating the tangential 
forces, normalized to the axisymmetric radial force field 
$Q_T(R)=F_T(R)/F_R(R)$.
In order to have a single measure for the bar strength, the maximum tangential
force, $Q_b$, in the bar region was used.
We also verified how well the ellipticity of a bar correlates with $Q_b$.

In the analysis we used JHK-images
of the 2 Micron All Sky Survey (2MASS) by constructing mosaics for most
of the galaxies. Some analytical force models were also carried out
to interpret our observational results, in which models bars were 
described by Ferrers-ellipsoids.
One of the most important results in this work is that favourable conditions 
for the onset of nuclear activity seem to be met in those early type galaxies, 
where the 
non-axisymmetric forces are rather small in the inner disks. 
\vskip 0.30cm

The main conclusions are the following:

\vskip 0.30cm

1. {\it The maximum ellipticity $\epsilon$ of the bar correlates 
quite well with the 
maximum tangential force $Q_b$ in the bar region.} This is a first clear
empirical indication showing that $\epsilon$ can be used as a mesure
of the bar strength. Based on our toy models
the small scatter in the diagram can be understood by means of 
different bar mass fractions in galaxies. Alternatively, the dispersion
can be partly due to the uncertainties in the vertical scale heights 
of the galactic disks.
\vskip 0.30cm

2. {\it The distance of the peak ellipticity, $r_{\epsilon}$, in the bar region, 
generally 
used as a measure of the bar length, systematically underestimates the 
length of the bar.} This was verified by comparing $r_{\epsilon}$ with 
$r_{bar}$, obtained by fourier analysis.  

\vskip 0.30cm

3. {\it In the active early type galaxies bars are on the average weaker 
(and their maximum ellipticities are
smaller) than in their non-active counterparts or in the late type
galaxies.} The KS-test shows that the probability that the samples of
active and non-active early type galaxies 
are drawn from similar populations is p=0.04
and 0.02 for $Q_b$ and $\epsilon$, respectively, the differences thus being
statistically significant. The mean $Q_b$ values
for the samples of active and non-active esrly type galaxies 
are $0.21 \pm 0.07$ and $0.35 \pm 0.19$.
\vskip 0.30cm

4. On the other hand, the scaled distances of the maximal tangential forces or
ellipticities, {\it $r_{Qb}/h$ and $r_{\epsilon}/h$ in the bar region, 
are \it largest for the 
active early type galaxies}. The differences in $r_{Qb}/h$ and 
$r_{\epsilon}/h$ between the active and
non-active early type galaxies are statistically significant
with p=0.0008 and p=0.007. The mean $r_{Qb}/h$ for the two subsamples
of early-type galaxies 
are $1.24 \pm 0.58$ and $0.51 \pm 0.28$, respectively. In comparison
to active early-type galaxies 
the values of these parameters are somewhat smaller for the active 
late type systems, and smallest for the non-active galaxies of any 
Hubble type. 
\vskip 0.30cm

5. We confirm the earlier result by Elmegreen $\&$ Elmegreen (1985), 
Martin (1995) and Regan $\&$ Elmegreen (1997) showing that bars are on the average 
longer for the early than for the late type galaxies. However, 
the dispersion especially for the early type galaxies is very 
large. 
\vskip 0.30cm

6. {\it Bar length $r_{bar}$ does not correlate with bar strength $Q_b$ for the
active early-type galaxies}, rather short bars
can have either weak, moderate or strong non-axisymmetric forces.
However, a weak correlation appears for the active late type galaxies 
and for all the non-active galaxies showing 
an increasing bar strength with the increasing bar length. 
\vskip 0.30cm

\section*{Acknowledgments}

This publication makes use of data products from the Two Micron All
Sky Survey, which is a joint project of the University of
Massachusetts and the Infrared Processing and Analysis
Center/California Institute of Technology, funded by the National
Aeronautics and Space Administration and the National Science
Foundation. It also uses the NASA/IPAC 
Extragalactic Database (NED), operated by the Jet Propulsion
Laboratory in Caltech. 
We acknowledge the foundations of Magnus Ehrnrooth, V\"ais\"al\"a
and Wihuri and the Academy of Finland of significant
financial support.

\vfill
\eject

\subsection{Figure captions}

{\bf Figure 1.} Comparison of the maximum ellipticities of bars 
($\epsilon=1-b/a_{min}$) and 
the locations of most elongated isophotes ($r_{\epsilon}$) between the 
measurements by Laine et al. (2001) 
and us, for the 6 galaxies common in our samples. Notice the very good 
agreement except for one galaxy (open symbol). For that galaxy the 
2MASS-image we use is not deep enough to reveal the maximum ellipticity.

{\bf Figure 2.} Bar strength $Q_b$ v.s. minor-to-major axis ratio
for the galaxies in our sample. The active and non-active early and 
late type
galaxies are shown separately with different symbols (same in all the 
subsequent figures).
In order to have our figure to be easily comparable with that by Block et al.
(2001), b/a was used instead of $\epsilon$.

{\bf Figure 3.} The histograms of the maximum tangential 
forces $Q_b$ and their distances $r_{Qb}/h$ for the barred active and
non-active galaxies are compared. The distances are scaled to the scale 
length of the disk, taken from Baggett et al. (1998) or measured
from the 2MASS-images by us.
In the $Q_b$-histogram one non-active galaxy has $Q_b$=1.4,
which was left out from the histogram. 

{\bf Figure 4.} The distance of the maximum non-axisym/-metric
force $r_{Qb}/h$ v.s. the maximum force $Q_b$, shown separately for 
the active and the
non-active early and late type barred galaxies in our sample. 

{\bf Figure 5.} The radial $Q_T$-profiles for the 
active
and the non-active early and late type barred galaxies. The radii are 
scaled to the 
scale length of the disk. The symbols denote the location of the assigned
maximum $Q_T$-values. 
Note that for some late type systems the $Q_T$-profiles
rise monotonically toward the center: in these cases the maximum was estimated
by eye eliminating the possible contribution of the artficial bulge
elongation in the deprojection. The uncertainty of these cases does not affect
our conclusions.

%{\bf Figure 6.} For the mixed-type and the non-barred galaxies, the comparison 
%of the
%frequences of the $Q_b$-values between the active and the non-active galaxies.

{\bf Figure 6.} In the upper frame the relation between the 
distances of 
the force maxima $r_{Qb}/h$ and the ellipticity maxima $r_{\epsilon}/h$
is shown, while in the lower frame $r_{\epsilon}$ is plotted against $r_{bar}$.
Notice that in the upper frame the symbols at $r_{Qb}$ = 10 correspond
to our adopted lower limit of $r_{Qb}$. Likewise for theree cases with
$r_{\epsilon}$=0 no definite value for $r_{\epsilon}$ could be obtained.

{\bf Figure 7.} In th eupper frame bar length $r_{bar}$ is plotted as a 
function of bar 
strength $Q_b$ for the  barred galaxies, and below a similar plot for the 
ellipticity measurements is shown.

{\bf Figure 8.}  Analytical toy models for the maximum ratio of tangential to
radial force ($Q_b$) and its location ($r_{Qb}/r_{bar}$), for
different bar axial ratios ($b/a$). 
As explained in more detail in the
text, the models consist of a bulge + disk combined with a Ferrers-bar.
Here $M_{bulge}/(M_{bar}+M_{disc})=0.3$, implying a short rising part of the
mean rotation curve. In the upper row $M_{bar}/(M_{bar}+M_{disc})=0.2, 0.4,
0.8$ are compared, while a fixed $n=1$ is assumed. In the lower row
$n=0,1,2$ while $M_{bar}/(M_{bar}+M_{disc})=0.4$ is fixed. The frames in
the left column show the rotation curves (thick curves) as well as the
radial $Q_T$-profiles for $b/a=0.5$ (thin lines). The middle column
displays the corresponding $Q_b$ vaues, while in the left $r_{Qb}/a$
are shown. Symbols denote our measurements.

{\bf Figure 9.} Same as Fig. 9, except that the bulge component is omitted,
leading to more rapidly rising central rotation curve.

{\bf Figure 10.} The absolute bar length $r_{bar}$ v.s. bar strength $Q_b$
for the active galaxies
showing separately Seyfert 2 and Seyfert 1-1.5 galaxies, LINERs and 
HII-galaxies. The mixed activity types are indicated by two superimposed 
symbols.

%\bye

%\epsffile{finalfig_fig1.eps}

\vfill
\eject
%\ifsinglecol
\vbox to 646pt{}
%\fi
\psfig{file=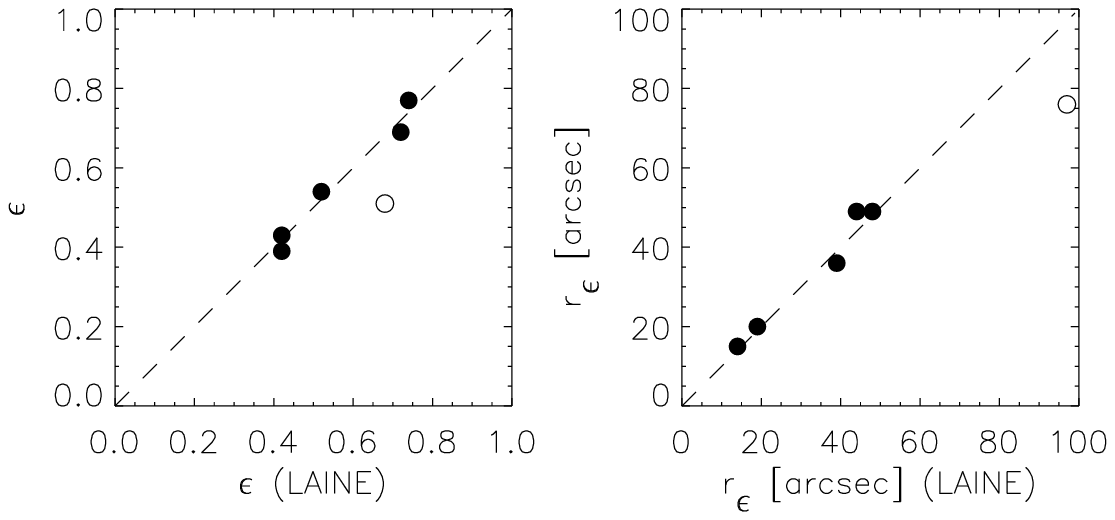,width=16cm}
Fig. 1
\vfill
\eject

\vbox to 646pt{}
\psfig{file=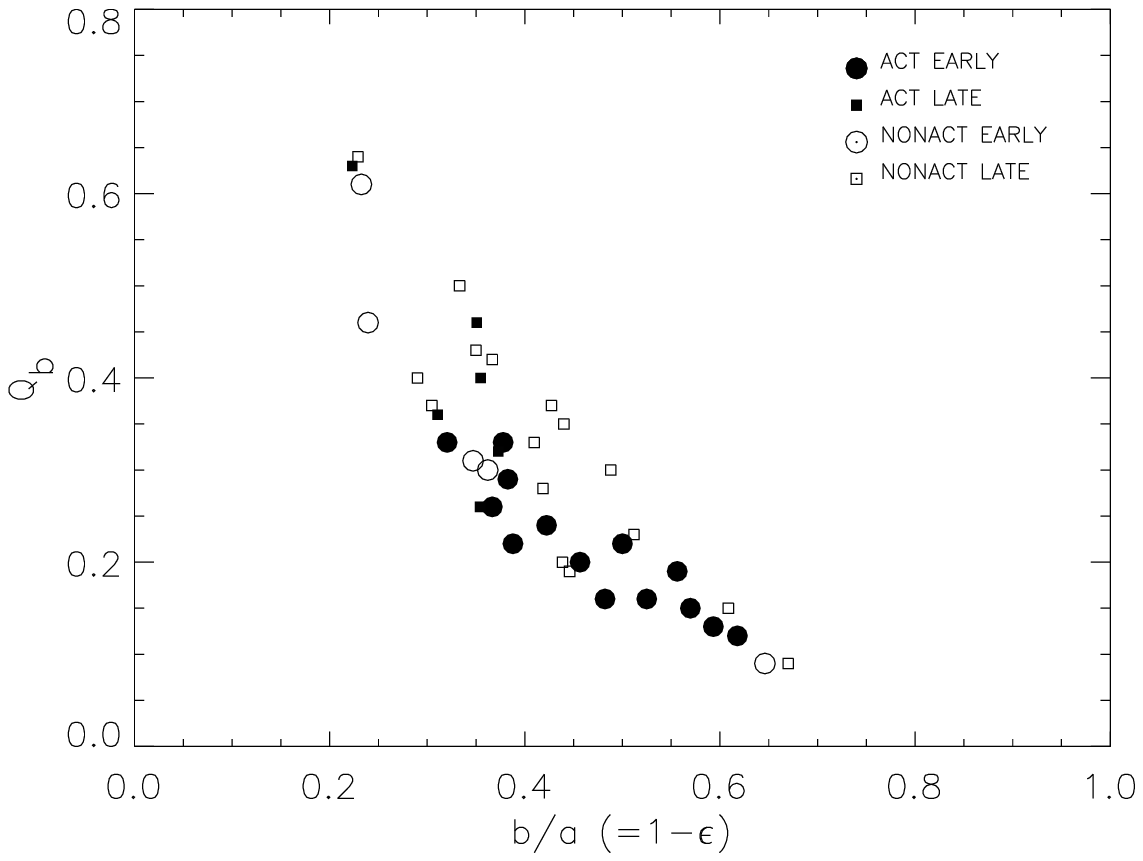,width=16cm}
Fig. 2
\vfill
\eject

 \vbox to 646pt{}
\psfig{file=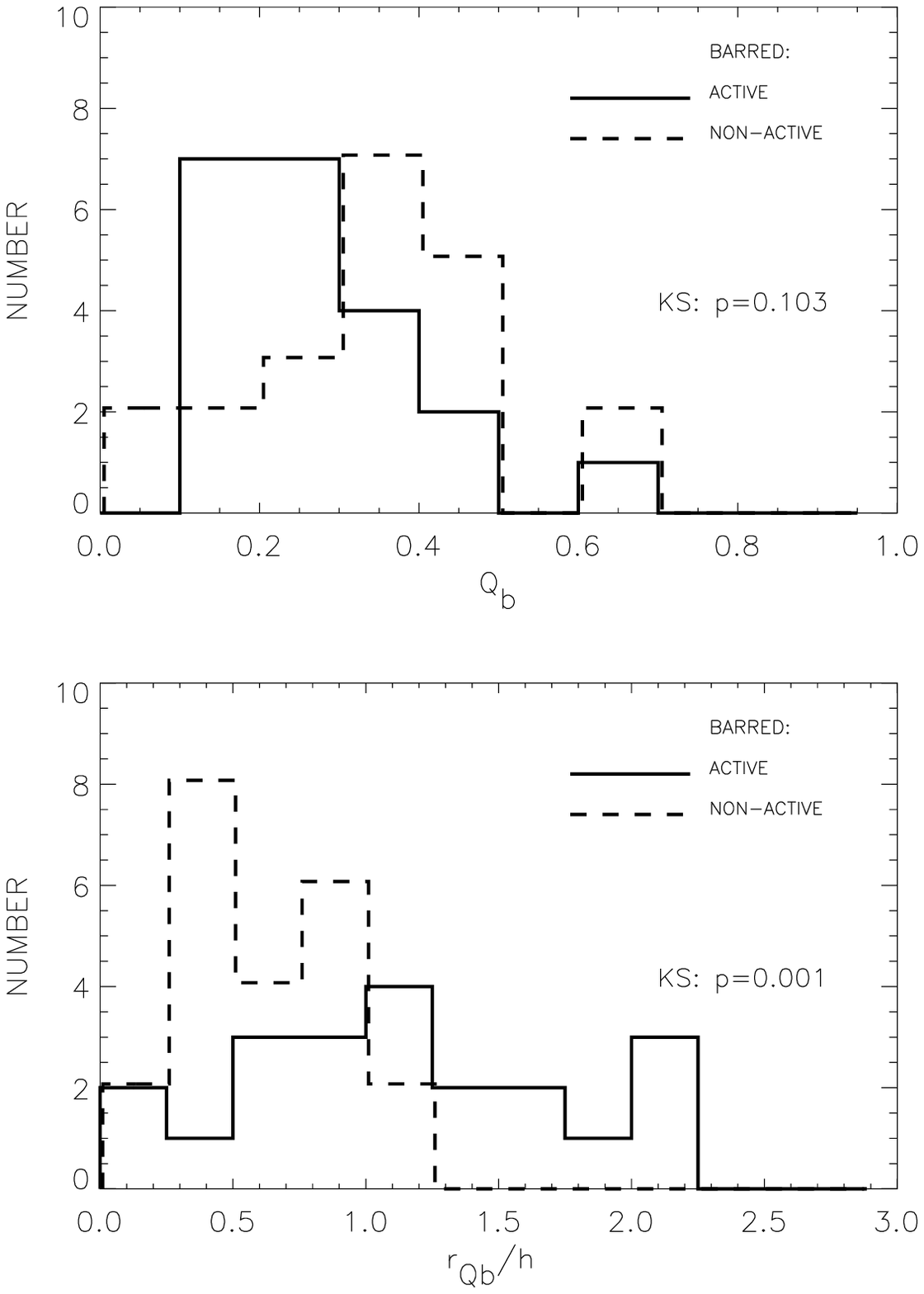,width=16cm}
Fig. 3
\vfill
\eject

 \vbox to 646pt{}
\psfig{file=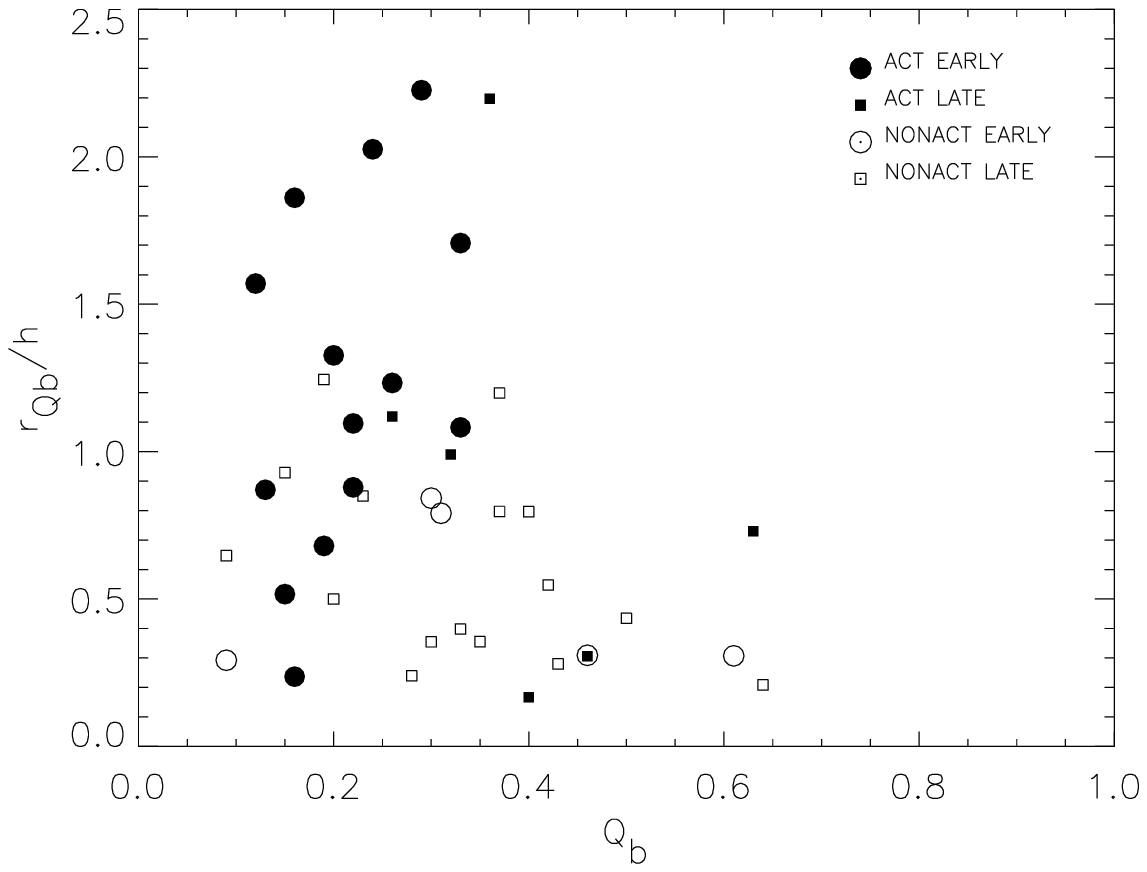,width=16cm}
Fig. 4
\vfill
\eject

 \vbox to 646pt{}
\psfig{file=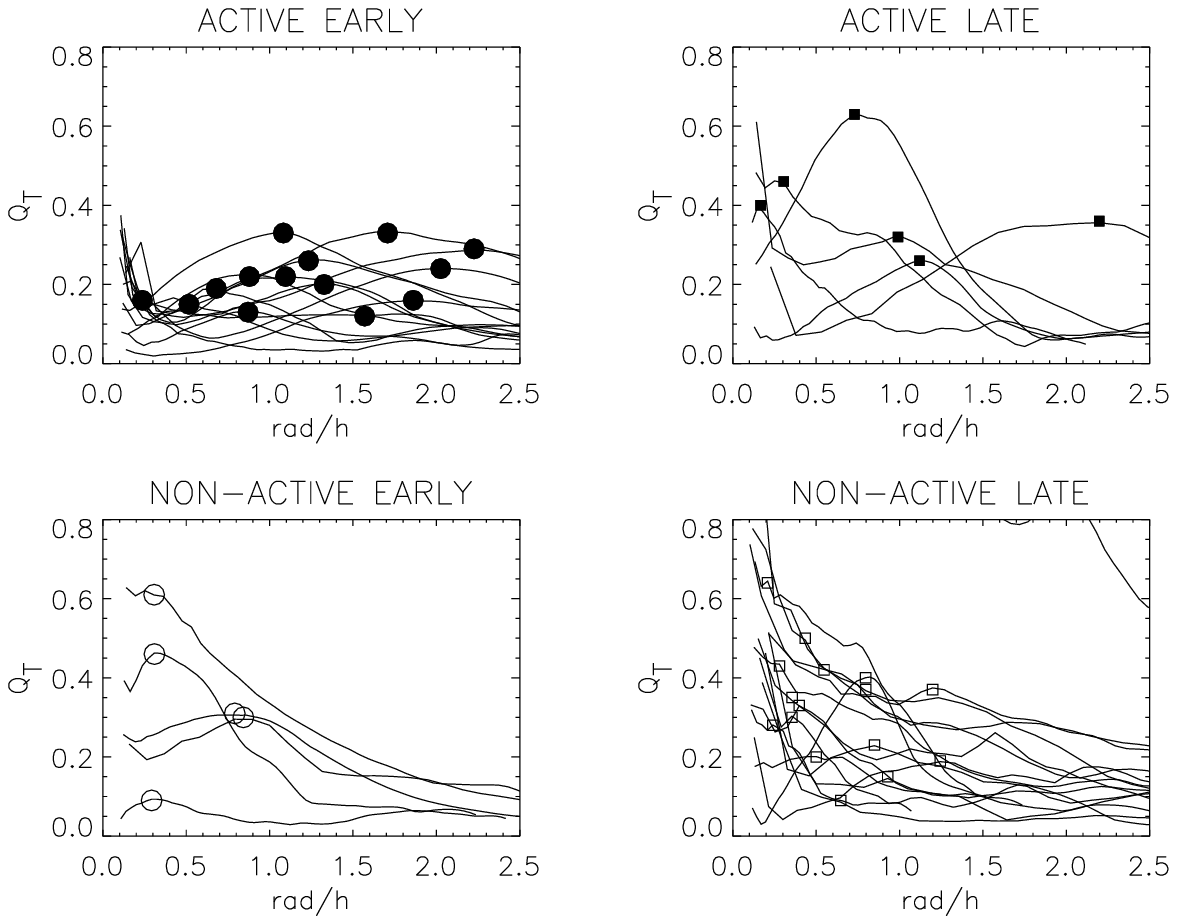,width=16cm}
Fig. 5
\vfill
\eject

 \vbox to 646pt{}
\psfig{file=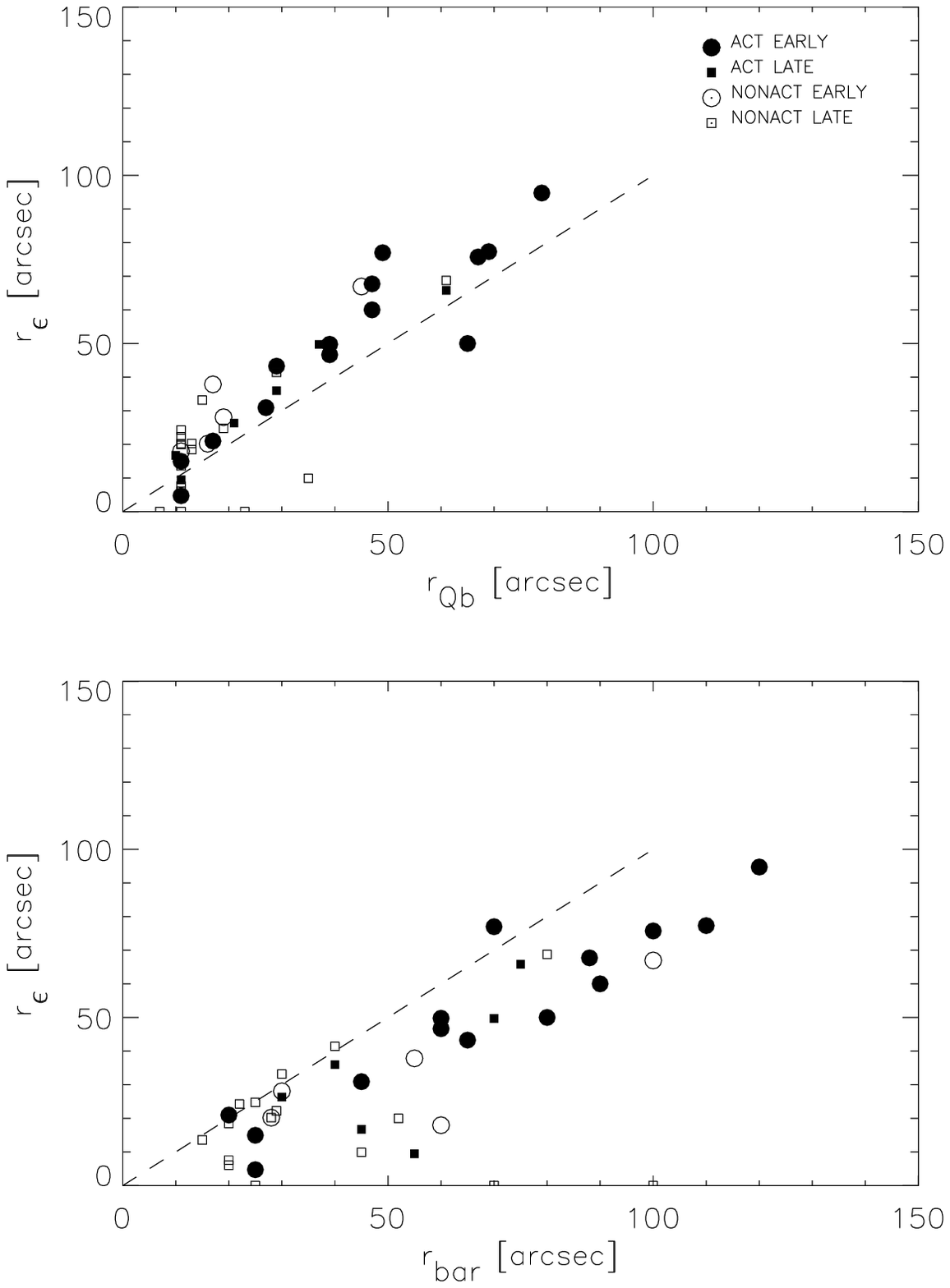,width=16cm}
Fig. 6
\vfill
\eject

 \vbox to 646pt{}
\psfig{file=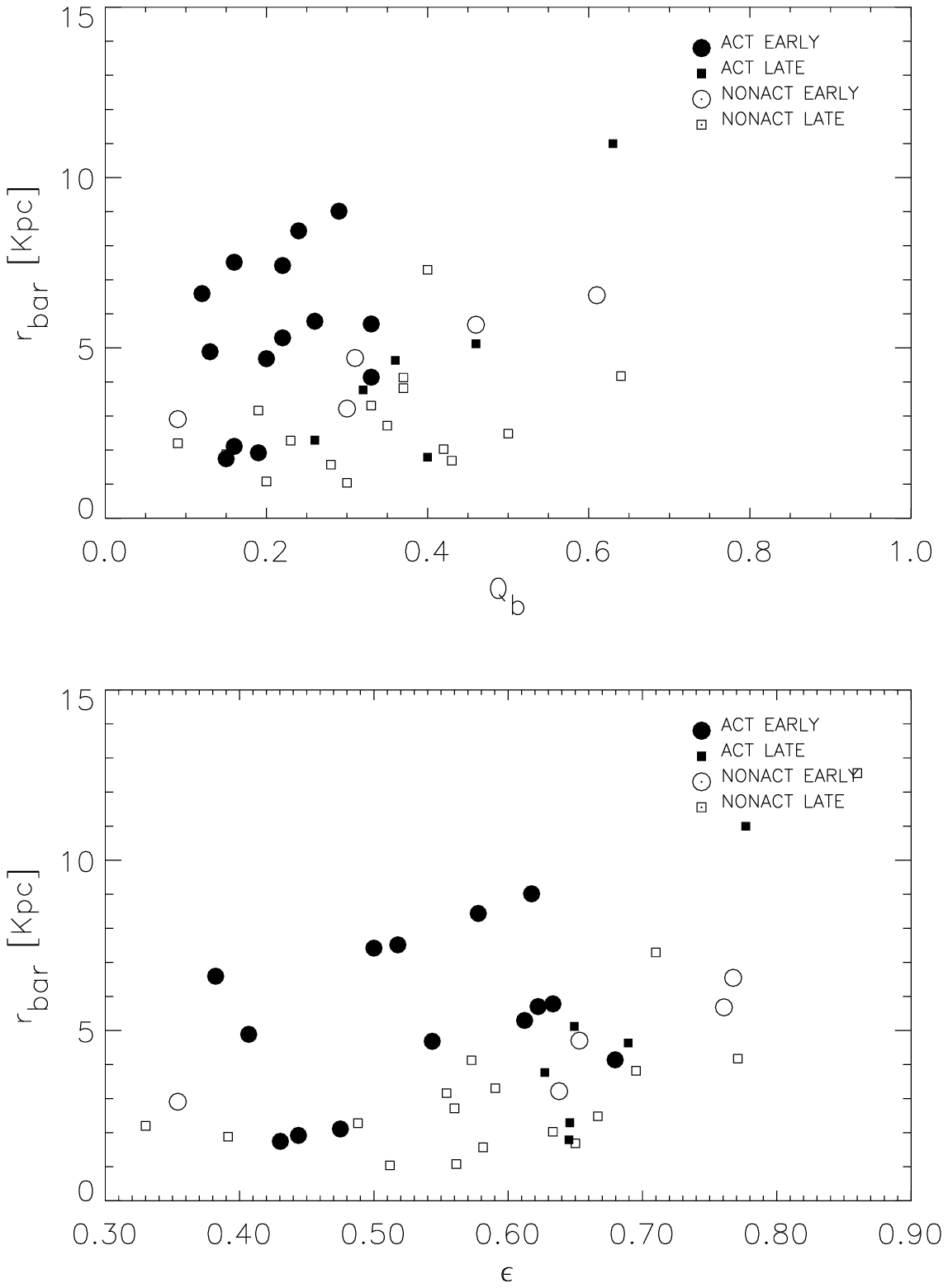,width=16cm}
Fig. 7
\vfill
\eject

 \vbox to 646pt{}
\psfig{file=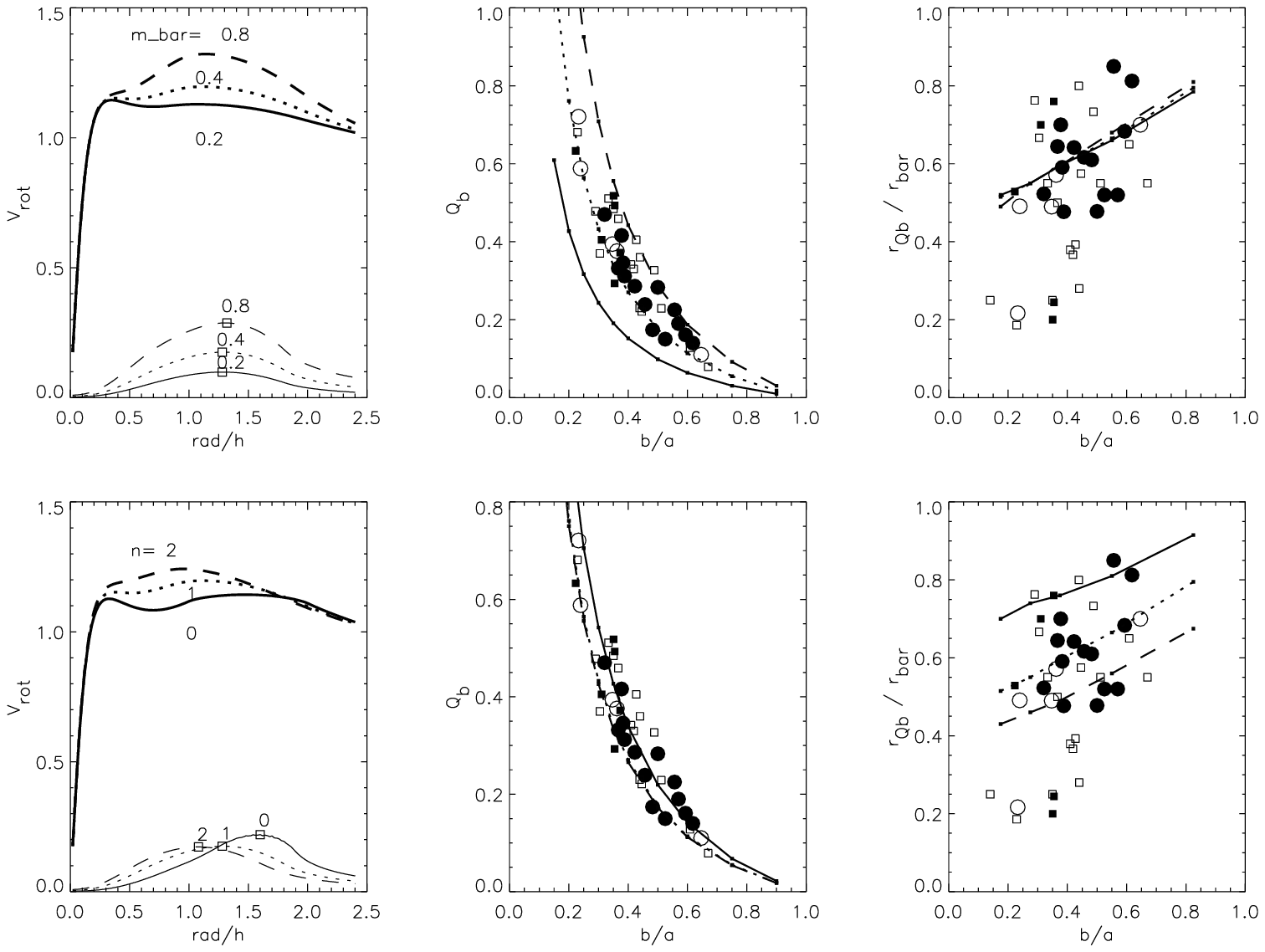,width=16cm}
Fig. 8
\vfill
\eject

 \vbox to 646pt{}
\psfig{file=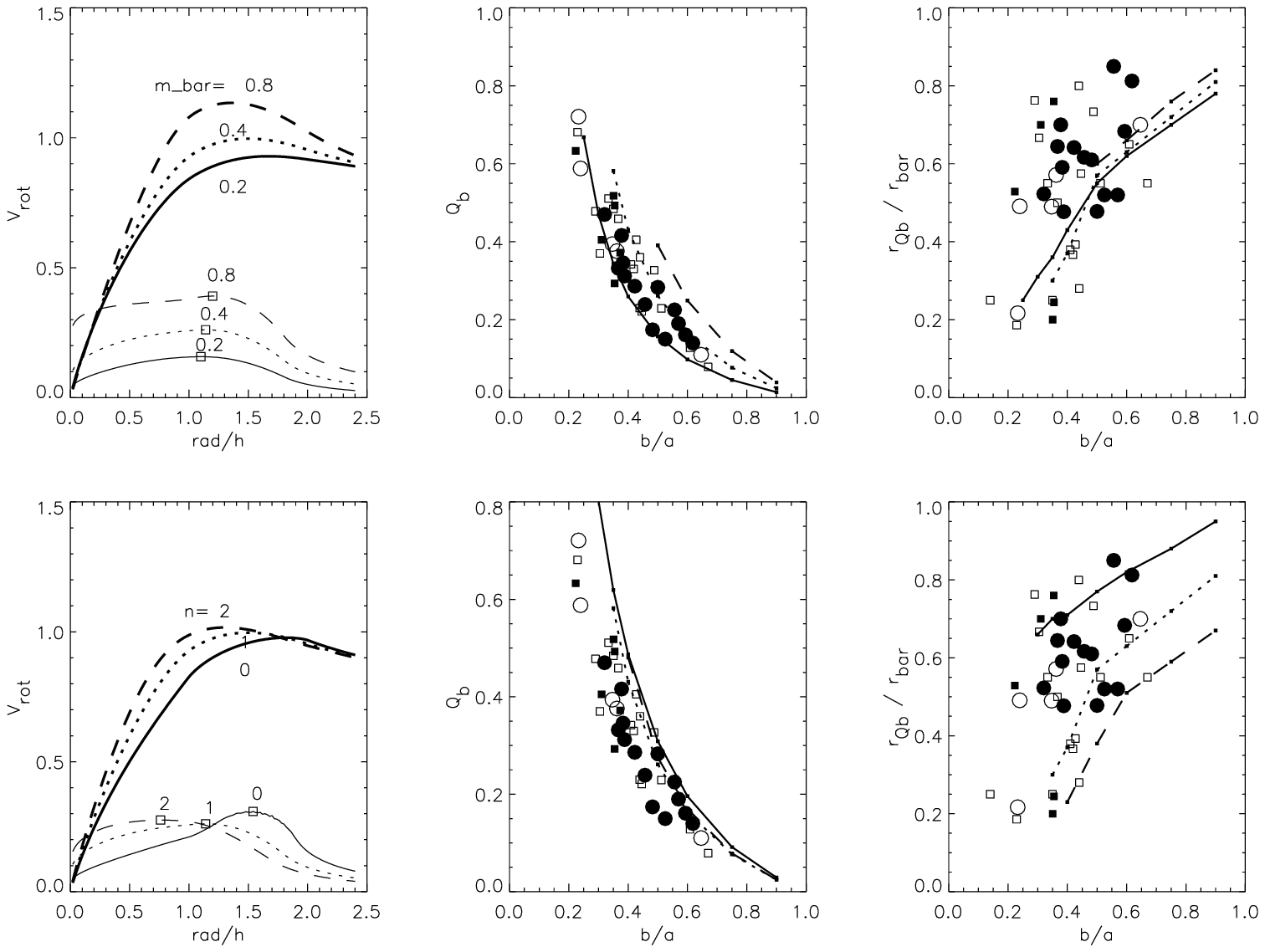,width=16cm}
Fig. 9
\vfill
\eject

 \vbox to 646pt{}
\psfig{file=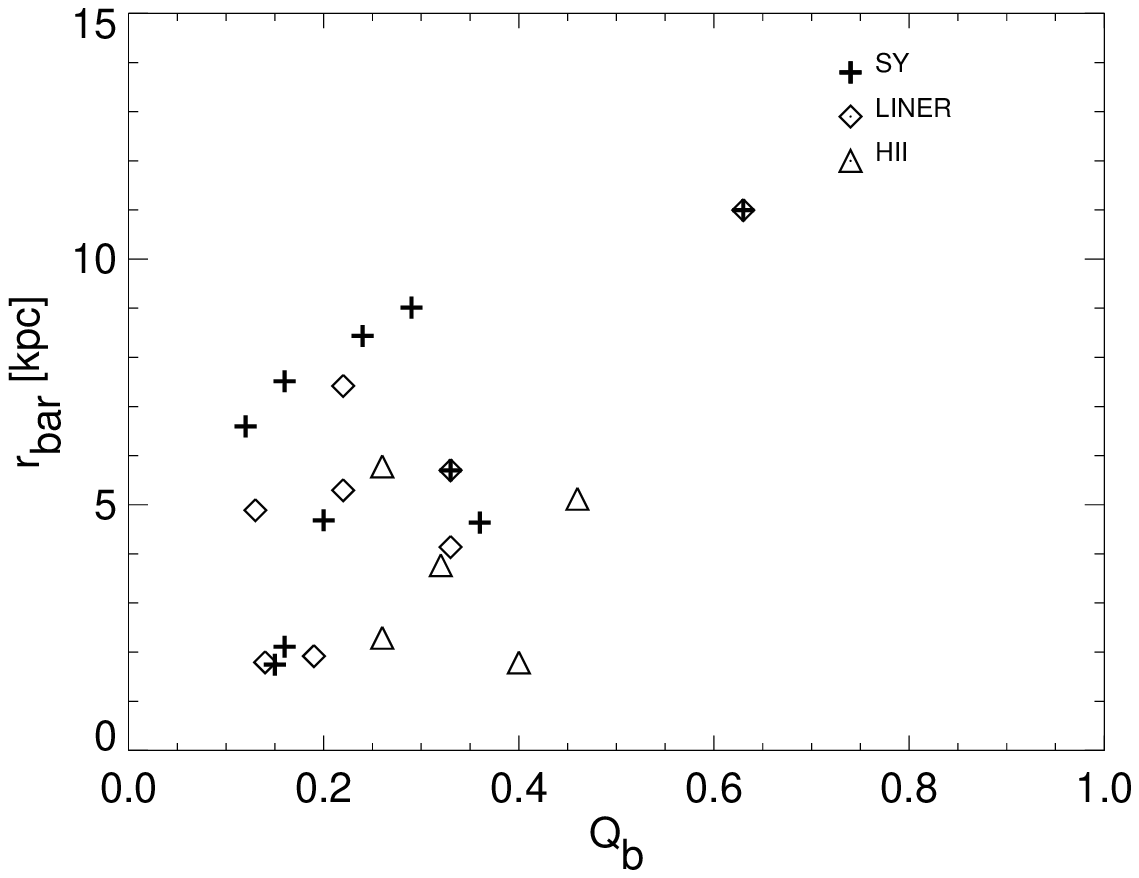,width=15cm}
Fig. 10
\vfill
\eject

\bye